\documentclass[twocolumn, aps, prb, showpacs, showkeys]{revtex4}
\usepackage{amsmath}
\usepackage{amssymb}
\usepackage{units}
\usepackage{graphicx}
\usepackage{psfrag}
\usepackage{tabularx}
\usepackage{booktabs}
\usepackage{bigstrut}
\usepackage[caption=false]{subfig}

\begin{document}

\title{How dielectric screening in two-dimensional crystals affects the convergence of excited-state calculations: Monolayer MoS$_2$}
\date{\today} \author{Falco H\"user}
\email{falco.hueser@fysik.dtu.dk} \affiliation{Center for Atomic-scale
  Materials Design (CAMD), Department of
  Physics\\
  Technical University of Denmark, 2800 Kgs. Lyngby, Denmark}
\author{Thomas Olsen} \affiliation{Center for Atomic-scale Materials
  Design (CAMD), Department of
  Physics\\
  Technical University of Denmark, 2800 Kgs. Lyngby, Denmark}
\author{Kristian S. Thygesen} \email{thygesen@fysik.dtu.dk}
\affiliation{Center for Atomic-scale Materials Design (CAMD),
  Department of
  Physics\\
  Technical University of Denmark, 2800 Kgs. Lyngby, Denmark}
\affiliation{Center for Nanostructured Graphene (CNG)\\
  Technical University of Denmark, 2800 Kgs. Lyngby, Denmark}

\begin{abstract}
We present first-principles many-body calculations of the dielectric constant,
quasiparticle band structure, and optical absorption spectrum of monolayer MoS$_2$
using a supercell approach. As the separation between the periodically repeated
layers is increased, the dielectric function of the layer develops a strong $q$ dependence
around $q=0$. This implies that denser $k$-point grids are required to converge the band gap
and exciton binding energies when large supercells are used. In the limit of infinite
layer separation, here obtained using a truncated Coulomb interaction,
a $45\times45$ $k$-point grid is needed to converge the G$_0$W$_0$ band gap and
exciton energy to within $\unit[0.1]{eV}$. We provide an extensive comparison with
previous studies and explain agreement and variations in the results.
It is demonstrated that too coarse $k$-point sampling and the interactions between the
repeated layers have opposite effects on the band gap and exciton energy,
leading to a fortuitous error cancellation in the previously published results.
\end{abstract}


\maketitle

\section{Introduction}
Atomically thin two-dimensional (2D) materials such as graphene,
hexagonal boron nitride, and transition metal dichalcogenides (TMDC)
possess unique electronic and optical properties including high
intrinsic carrier mobilities,
\cite{Novoselov_PNASU2005, Radis_Nature2011, Kaasbjerg_PRB2012}
tunable band gaps,\cite{Shi_PRB2013, Yan_PRB2012} and strong light-matter
interactions.\cite{Britnell_Science2013, Mak_PRL2010,
Splendiani_NanoLett2010, Bernardi_NanoLett}
These features, combined with the possibility of engineering their
electronic properties further via strain, alloying or stacking,
make the 2D materials ideal as building
blocks for new opto-electronic structures and devices with minimal
sizes and performances surpassing present technologies.

After the intense focus on graphene, the TMDCs are now attracting
increasing interest.\cite{Wang_Nature2012} This stems mainly from
the greater variation in their electronic properties including both 
semiconducting and metallic behavior. So far, the most intensively
studied single-layer TMDC is the semiconductor MoS$_2$.
Nanostructured forms of MoS$_2$ have previously
been explored as potential catalysts for desulferization of crude oil
and more recently for (photo)-electrochemical hydrogen evolution.
\cite{Bollinger_PRL2001, Jaramillo_Science2007, Zhong_JACS2008}
Bulk MoS$_2$ is composed of two-dimensional sheets held together
by weak van der Waals forces and individual sheets can be isolated
by exfoliation techniques similar to those used to produce graphene.
\cite{Novoselov_PNASU2005} Single layers of MoS$_2$ therefore comprise
highly interesting two-dimensional systems with a finite band gap and
have recently been proposed for nano-electronics applications.
\cite{Radis_Nature2011}

The optical properties of MoS$_2$ have been thoroughly studied
experimentally.\cite{Frindt_PRSLA1963, Evans_PRSLA1965, Wilson_AdvPhys1969, 
Neville_PSSB1976, Frey_PRB1998, Wilcoxon_JAP1997}
The absorption spectrum shows two distinct low energy peaks at
$\unit[1.88]{eV}$ and $\unit[2.06]{eV}$, which are denoted by A and B,
respectively,\cite{Coehoorn_PRB1987} and derive from direct transitions between
a split valence band and the conduction band at the K point of the Brillouin
zone. Their Rydberg satellites, Zeeman splitting, and dependence on crystal
thickness have been investigated in detail.\cite{Neville_PSSB1976} Recently,
the quantum yield of luminescence from MoS$_2$ was shown to increase
dramatically when the sample thickness was changed from a few layers to a
monolayer\cite{Mak_PRL2010, Splendiani_NanoLett2010} indicating a transition
to a direct band gap in the single layer.

In the past couple of years a number of theoretical studies of the
electronic band structure and optical excitations in monolayer MoS$_2$
have been published.\cite{Shi_PRB2013, Chei_PRB2012, Komsa_PRB2012, 
Ataca_JPCC2011, Rama_PRB2012, Molina_PRB2013, Conley_NanoLett2013}
These studies are based on many-body perturbation theory in the GW
approximation (mainly the non-selfconsistent G$_0$W$_0$ approach) for
the band structure and the Bethe-Salpeter equation (BSE) with a
statically screened electron-hole interaction for the optical
excitations. As is standard practice, the calculations have been
performed on a supercell geometry where the MoS$_2$ layers have been
separated by $\unit[10]{}-\unit[20]{\text{\AA}}$ vacuum and the Brillouin Zone
(BZ) sampled on grids ranging from 
$6\times 6$ to $15\times 15$. With these parameters G$_0$W$_0$ band
gaps in the range $\unit[2.6]{}-\unit[3.0]{eV}$, and G$_0$W$_0$-BSE exciton
binding energies of $\unit[0.6]{}-\unit[1.1]{eV}$, have been reported.
Moreover, both direct\cite{Ataca_JPCC2011, Chei_PRB2012, Komsa_PRB2012,
Rama_PRB2012, Molina_PRB2013} and indirect\cite{Shi_PRB2013} band gaps
have been found at the G$_0$W$_0$ level, while only direct gaps have been
obtained with self-consistent GW\cite{Chei_PRB2012} and GW$_0$.
\cite{Shi_PRB2013, Conley_NanoLett2013}
When comparing these values, it should be kept in mind that both size
and nature of the band gap of MoS$_2$ depends
sensitively on the in-plane lattice parameter, $a$.\cite{Shi_PRB2013}

One of the most fundamental quantities describing the electronic
structure of a material is the dielectric function. The dielectric
properties of atomically thin 2D materials are quite different from
their 3D counterparts.\cite{Keldysh_JETP1979} For example plasmons
in 2D metals have acoustic  dispersion relations
($\omega_p(q)\to 0$ as $q\to 0$), and screening is 
generally much weaker leading to strong exciton binding energies in 2D 
semiconductors. Reported static dielectric constants
for monolayer MoS$_2$ obtained using the supercell approach lie in the
range $\unit[4.2]{}-\unit[7.6]{}$ (for in-plane polarization).
\cite{Chei_PRB2012, Rama_PRB2012, Wirtz_PRB2011} These values have 
been used to rationalize the exciton binding energy in MoS$_2$
using the simple Mott-Wannier model.

In this paper, we present an in-depth study of the dielectric
function, quasiparticle band structure and excitonic states in
monolayer MoS$_2$. We focus on separating the spurious interlayer
screening from the intrinsic intralayer screening in supercell
geometries, and the consequences that the physics of screening in 2D
has for the convergence of many-body excited state calculations.
The 3D macroscopic dielectric constant, as used for solids, converges
to 1 for all $q$ vectors in the limit of infinite separation of the
layers and is thus meaningless for a 2D material. We use an alternative
approach to calculate the dielectric constant by averaging the total
field over the material rather than the supercell. This
2D dielectric constant shows strong $q$-dependence for small wave
vectors and becomes exactly 1 for $q=0$.  This property has important
consequences for the $k$-point convergence of many-body calculations.

In general, the use of a truncated Coulomb interaction is essential to
avoid interlayer screening which decays slowly with the layer
separation, $L$. The interlayer screening yields too
large dielectric constant for wave vectors $q<1/L$.
As a consequence, the G$_0$W$_0$ band gaps and exciton energies are
$\unit[0.5]{eV}$ too low on average for layer separations of around
$\unit[20]{\text{\AA}}$. For larger layer separations, the strong
$q$-dependence of the dielectric constant for small $q$ implies that a
$k$-point grid of at least $45\times 45$ is required to converge band gaps and
exciton energies to $\unit[0.1]{eV}$. For $k$-point grids below
$15\times 15$ the band gap is at least $\unit[0.5]{eV}$ too large in the limit
$L \to \infty$. Thus the effect of interlayer screening and
too coarse $k$-point grids partially cancel out leading to reasonable
values for the band gap and exction binding energy with 
underconverged parameters as applied in previous studies.

The paper is organized as follows. In Sec. \ref{sec:band} we present
G$_0$W$_0$ band structures and study the convergence of the gap with respect to
interlayer separation and $k$-point sampling. In Sec. \ref{sec:epsilon} we
show calculations for the 2D dielectric constant and explain the
origin of the slow $k$-point convergence of the band gap. In Sec. \ref{sec:BSE}
we present many-body calculations of the lowest excitons and analyse their
convergence with layer separation and $k$-point sampling. Our conclusions are
given in Sec. \ref{sec:conclusion}.

\section{Quasiparticle band structure}\label{sec:band}
In this section we demonstrate that GW band structures for monolayer MoS$_2$
converge extremely slow with respect to the interlayer separation.
In order to obtain well converged results (within $\unit[0.1]{eV}$), the use
of a truncated Coulomb interaction is inevitable, along with a $k$-point grid
of around $45\times 45$. Previously reported calculations with the full
Coulomb interaction have employed only separation between
$\unit[10]{}$ and $\unit[20]{\text{\AA}}$ and used from
$6\times 6$ to $12\times 12$ $k$ points.
The resulting band structures are, however, somewhat saved by a
fortunate error cancellation between the two effects.
 
\subsection{Computational details}\label{subsec:GWdetails}
All our calculations have been performed with the projector augmented
wave method code GPAW.\cite{GPAW}\footnote{The \texttt{gpaw} code is available
as  a part of the CAMPOS software: \texttt{www.camd.dtu.dk/Software}}
The Kohn-Sham  wave functions and energies of monolayer MoS$_2$ were calculated
in the local density approximation (LDA) using a plane wave basis with
cut-off energy $\unit[400]{eV}$. The $4s$ and $4p$ semicore electrons of Mo
were explicitly included in all calculations.  Unless otherwise stated
the calculations have been performed for the experimental lattice
constant of $\unit[3.16]{\text{\AA}}$. One-shot G$_0$W$_0$
calculations were performed using the LDA wave functions and
eigenvalues to obtain the G$_0$W$_0$@LDA quasiparticle energies. A
plane wave cut off of $\unit[50]{eV}$ and 200 bands were used for
the dielectric function, screened interaction and GW self-energy.
Convergence with respect to these parameters has been checked very carefully.
With these values band gaps were found to be converged within
around $\unit[10]{meV}$. The plasmon pole approximation for
the dielectric function was found to yield QP energies within $\unit[0.1]{eV}$
of those obtained from full frequency dependence and was consequently
used in all calculations. To avoid interaction between the
periodically repeated MoS$_2$ sheets, we have applied a truncated
Coulomb interaction of the form $v_c(\mathbf r)=(1/r)\theta(R_c-z)$,
following Refs. \onlinecite{Rozzi_PRB2006} and
\onlinecite{Ismail-Beigi_PRB2006}. For details on the implementation of the GW
method in the GPAW code we refer to Ref. \onlinecite{Hueser_PRB2013}.
We note that we have used a numerical averaging of the head of the
screened potential $W_{\mathbf{0}\mathbf{0}}(\mathbf{q})$ for all
wavevectors $\mathbf{q}$ in the Brillouin zone (similar to Eq. \ref{eq:W_int}
in Sec. \ref{subsec:screenedinteraction}). This was found to be
crucial in all calculations with the full $1/r$ Coulomb interaction.

\begin{figure}[t]
\begin{centering}
\includegraphics[width=\columnwidth,clip=]{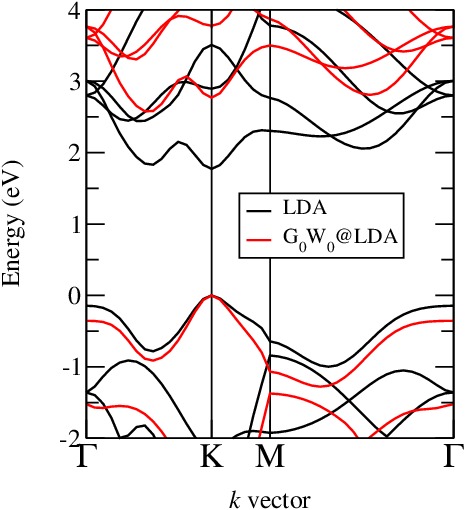}
\par\end{centering}
\caption{\label{fig:bandstructure}(Color online). Bandstructure of monolayer 
MoS$_2$ calculated with LDA and G$_0$W$_0$@LDA using $45\times 45$ $k$ points
and a truncated Coulomb interaction to avoid interaction between periodically 
repeated layers. The valence band tops have been aligned.}
\end{figure}

\begin{figure}[t]
\begin{centering}
\includegraphics[width=0.8\columnwidth,clip=]{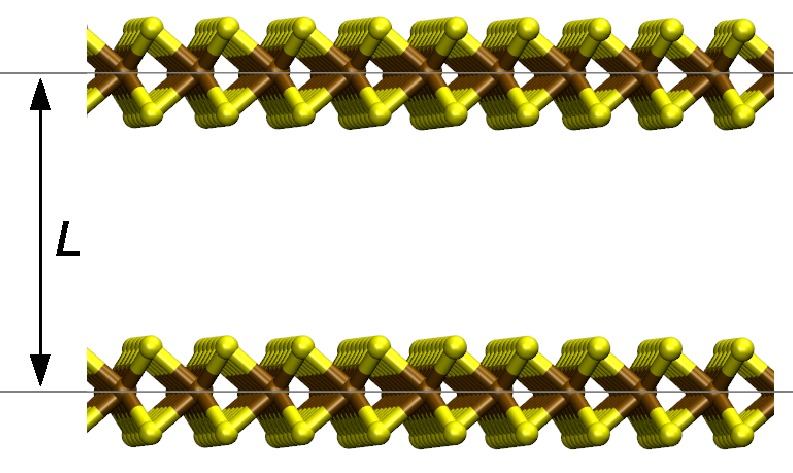}
\par\end{centering}
\caption{\label{fig:distance}Definition of the interlayer separation, $L$.}
\end{figure}

\begin{figure}[t]
\begin{centering}
\includegraphics[width=0.85\columnwidth,clip=]{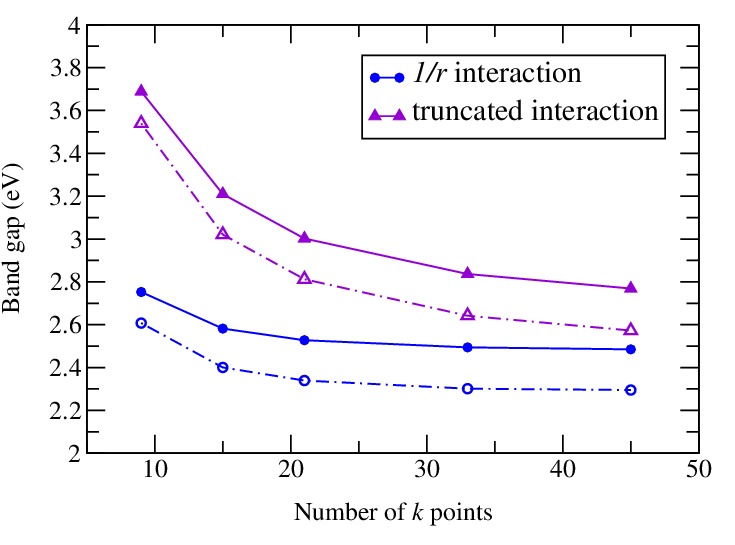}
\par\end{centering}
\caption{\label{fig:kpoints}(Color online). Direct (full symbols) and indirect
(open symbols) G$_0$W$_0$ band gaps as function of the number of $k$ points
in one of the in-plane directions for a layer separation of
$L=\unit[23]{\text{\AA}}$.}
\end{figure}

\begin{figure}[t]
\begin{centering}
\includegraphics[width=\columnwidth,clip=]{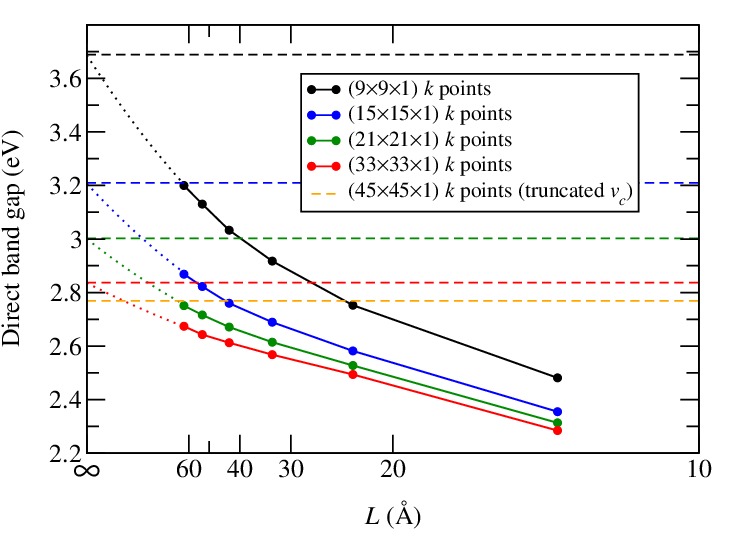}
\par\end{centering}
\caption{\label{fig:GWgaps}(Color online). Direct G$_0$W$_0$ band gap
plotted as a function of interlayer distance for different $k$-point samplings
with the full $1/r$ interaction. Dotted lines serve as a guide for the eye
to extrapolate for $L \to \infty$. They were obtained by fitting all values for
$L > \unit[30]{\text{\AA}}$, including the results with the Coulomb truncation,
to a quadratic function. Dashed horizontal lines
indicate the calculated values with the truncated Coulomb interaction.}
\end{figure}

\begin{figure}[t]
\begin{centering}
\includegraphics[width=\columnwidth,clip=]{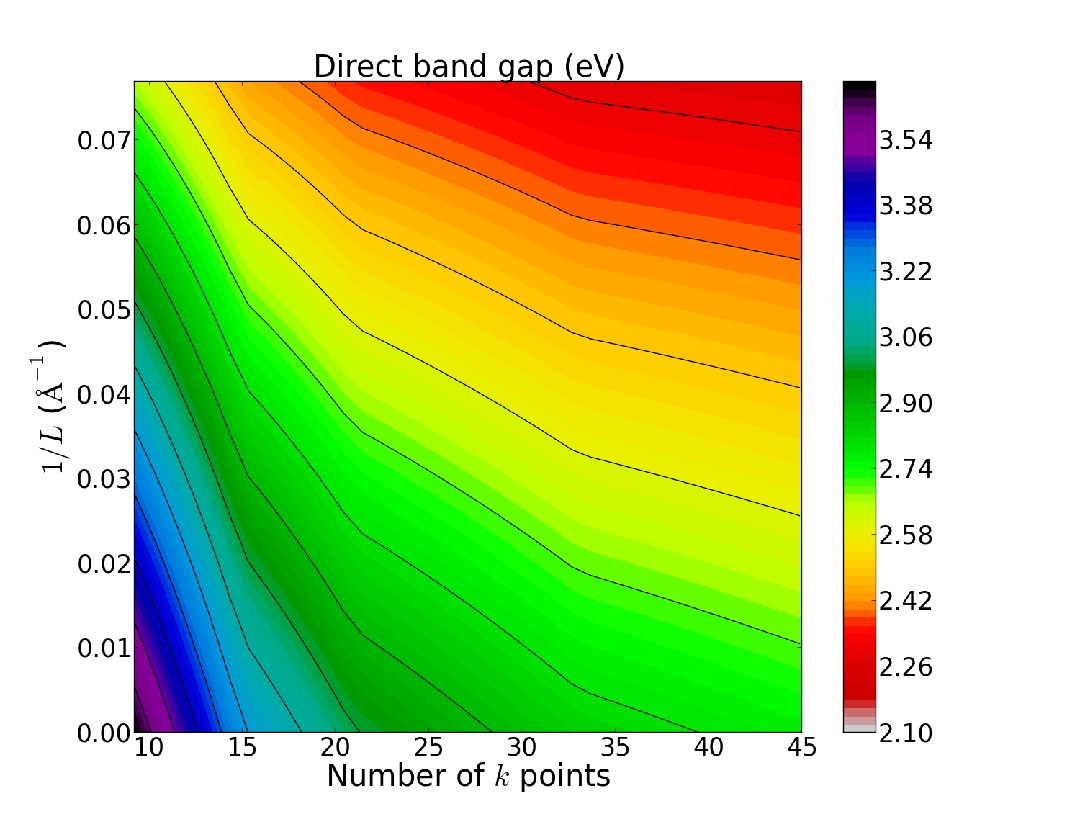}
\par\end{centering}
\caption{\label{fig:GWcontourplot}(Color online). Contour plot of the direct
G$_0$W$_0$ band gap as a function of the inverse interlayer distance and number
of $k$ points in one of the in-plane directions with the full $1/r$
interaction. Contour lines are separated by $\unit[0.1]{eV}$.
Interpolation from splines was used.}
\end{figure}

\subsection{Results}\label{subsec:GWresults}
The band structure calculated using $45\times 45$ $k$ points and the
truncated Coulomb interaction is shown in Fig.
\ref{fig:bandstructure}. At the LDA level, we find a direct band gap
at the K point of $\unit[1.77]{eV}$ while the smallest indirect gap of
$\unit[1.83]{eV}$ occurs from $\Gamma$ to a point along the $\Gamma$-K
direction. In contrast, G$_0$W$_0$ predicts an indirect gap of
$\unit[2.58]{eV}$ and at direct gap at K of $\unit[2.77]{eV}$.

In Fig. \ref{fig:kpoints} we show the convergence of both the
direct and the indirect band gap with respect to the $k$-point grid
for a fixed interlayer separation of $\unit[23]{\text{\AA}}$
(see Fig. \ref{fig:distance} for the definition of $L$).
It is clear that a very dense $k$-point grid
is needed in order to obtain well-converged results with the truncated
Coulomb interaction. For $45\times 45$ $k$ points, band gaps are
converged within less than $\unit[0.1]{eV}$, while this is already
the case for $15\times 15$ $k$ points with the bare Coulomb interaction.
However, the calculated values are too low.
The slow convergence with respect to $k$ points when the truncation is
used will be discussed in detail in Sec. \ref{subsec:screenedinteraction}.

We see that results do not converge independently with respect to the
number of $k$ points and the interlayer separation. In Fig. \ref{fig:GWgaps},
we plot the $L$-dependence of the direct band gap for different
$k$-point samplings with the bare interaction. The $k$-point dependence
becomes much stronger for large $L$. For $L \to \infty$, the values are
expected to converge to the results calculated with the truncation
(indicated by dotted lines). They seem to exhibit a linear $1/L$ behaviour
only for $L >  \unit[50]{\text{\AA}}$ or very dense $k$-point samplings.
Fig. \ref{fig:GWcontourplot} shows all results and interpolated values
in a contourplot as a function of $1/L$ and number of $k$ points.
The effects of using more $k$ points and increasing $L$ are of different sign
and partially cancel each other. This is the reason, why different choices
of the two parameters yield the same results. Especially, the band gaps
calculated with $9\times 9$ $k$ points and $L = \unit[23]{\text{\AA}}$ and
$15\times 15$ $k$ points and $L = \unit[43]{\text{\AA}}$ are the same
as with $45\times 45$ $k$ points and infinite $L$.
This seems, however, conincidental and we do not expect it to be the
case for other systems.

We note that all calculations have been performed with a single $k$ point in
the direction perpendicular to the layer. This is, however, insufficient
for small interlayer distances. For $L = \unit[13]{\text{\AA}}$, we find
an increase of the band gap of around $\unit[0.2]{}$ - $\unit[0.3]{eV}$
when at least 3 $k$ points are used, for example.
For $L > \unit[20]{\text{\AA}}$ or use of the truncation, this effect is
negligible.

\subsection{Comparison with previous work}\label{subsec:GWliterature}
In table \ref{tab:gaps} we show our converged results obtained with
the truncated Coulomb interaction and $45\times 45$ $k$ points together
with previous G$_0$W$_0$ results from the literature. For each
reference we show the values used for the lattice constant, the
interlayer separation and the $k$-point sampling. It can be seen that all the
previous calculations have used small layer separations and no truncation
method. As pointed out in the preceding discussion, this gives a fast
$k$-point convergence. A properly converged calculation, however, requires
larger separations and thereby more $k$ points. But as a consequence
of a cancellation of errors, a calculation with $\unit[19]{\text{\AA}}$
layer separation and $12\times 12$ $k$ points yields almost the same band
gaps as our converged result (within $\unit[0.15]{eV}$).
We are thus led to conclude that the reasonable agreement between our results
and previous ones are to a large extent fortuitous.

Furthermore, the effect of strain can have a large impact on the
MoS$_2$ band gap. As demonstrated in Ref. \onlinecite{Shi_PRB2013},
using $12\times 12$ $k$ points and $\unit[19]{\text{\AA}}$ layer
separation, the G$_0$W$_0$ band gap for the experimental lattice
constant of $\unit[3.160]{\text{\AA}}$ is indirect. 
With a lattice constant of $\unit[3.190]{\text{\AA}}$,
corresponding to $1 \%$ strain, the gap changes to direct.
The lowering of the direct band gap becomes even  more pronounced for
larger lattice constants. As can be seen from the table  our converged
results predict the same trend, in particular the decrease of the
direct gap as function of strain, with our values for the direct gap being 
generally $\unit[0.2]{eV}$ larger. We note that for $\unit[3.255]{\text{\AA}}$,
the smallest indirect transition occurs from the $\Gamma$ point at the valence
band to the K point at the conduction band. This is also in agreement with
Ref. \onlinecite{Shi_PRB2013}. In the partially self-consistent GW$_0$
calculations of Ref. \onlinecite{Conley_NanoLett2013}, the opposite trend was
found, namely a transition from a direct to indirect band gap for $\sim 5$
\% strain. However, a layer separation of only $\unit[12]{\text{\AA}}$
and less than 9 $k$ points in the in-plane directions were used.

\begin{table*}[t]
\begin{centering}
\caption{\label{tab:gaps}Calculated G$_0$W$_0$ band gaps obtained in present 
work and compared with previous results from the literature. All our 
calculations have been performed using a truncated Coulomb interaction.}
\begin{tabularx}{\textwidth}{l @{\hspace{0.5cm}} | c @{\hspace{1cm}} c 
@{\hspace{1cm}} c @{\hspace{1cm}} c @{\hspace{1cm}} r @{\hspace{0.5cm}} r 
@{\hspace{0.5cm}}}
\hline\hline
& & & & & \multicolumn{2}{c}{$E_{\text{gap}}$ (eV)} \bigstrut\\
\cline{6-7}
Ref. & starting point & $a$ (\AA) & number of $k$-points
& layer separation (\AA) & direct & indirect \bigstrut\\
\hline
This work 		   	& LDA & 3.16  & $45\times45\times1$ &
23 (truncated $v_c$) & 2.77 	   & 2.58        \\
This work 		   	& LDA & 3.19  & $45\times45\times1$ &
23 (truncated $v_c$) & 2.65 	   & 2.57        \\
This work 		   	& LDA & 3.255 & $45\times45\times1$ &
23 (truncated $v_c$) & 2.41 	   & 2.51        \\
Ref. \onlinecite{Molina_PRB2013}& LDA & 3.15  & $18\times18\times1$ &
24                   & 2.41 	   & $\sim 2.40$ \\
Ref. \onlinecite{Shi_PRB2013}   & PBE & 3.16  & $12\times12\times1$ &
19 		     & $\sim$ 2.60 & 2.49 	 \\
Ref. \onlinecite{Shi_PRB2013}   & PBE & 3.19  & $12\times12\times1$ &
19 		     & 2.50 	   & $\sim$ 2.55 \\
Ref. \onlinecite{Shi_PRB2013}   & PBE & 3.255 & $12\times12\times1$ &
19 		     & 2.19 	   & 2.19        \\
Ref. \onlinecite{Chei_PRB2012}  & LDA & 3.16  & $8\times8\times2$   &
19 		     & 2.96 	   & --          \\
Ref. \onlinecite{Komsa_PRB2012} & PBE & 3.18  & $12\times12\times1$ &
20+$1/L$ extrapolation& 2.97       & 3.26 	 \\
Ref. \onlinecite{Komsa_PRB2012} & PBE & 3.18  & $12\times12\times1$ &
20		     & $\sim$ 2.60 & $\sim$ 2.85 \\
Ref. \onlinecite{Ataca_JPCC2011}& LDA & 3.11  & $12\times12\times1$ &
13		     & 2.57        & --          \\
Ref. \onlinecite{Rama_PRB2012}  & HSE & 3.18  & $6\times6\times1$   &
15                   & 2.82        & $\sim$ 3.00 \\
Ref. \onlinecite{Ding_PhysicaB2011} & PBE & 3.19&$15\times15\times1$&
15                   & 2.66        & --          \\
\hline\hline
\end{tabularx}
\end{centering}
\end{table*}

In Ref. \onlinecite{Komsa_PRB2012}, the band gap was determined by extrapolating
from $L = \unit[20]{\text{\AA}}$ to infinite layer separation,
under the assumption that the gap scales linearly with the inverse distance 
between the layers. The obtained values for the direct
and indirect band gaps are $\sim\unit[3.0]{}$ and $\sim\unit[3.3]{eV}$,
respectively. This is consistent with our findings using the truncated
Coulomb interaction, the same lattice constant of $\unit[3.18]{\text{\AA}}$
and the same (under-converged) $k$-point grid of $12\times 12$ as in Ref. 
\onlinecite{Komsa_PRB2012}.

From our studies, we conclude that the G$_0$W$_0$@LDA band gap of monolayer 
MoS$_2$ is indirect with a value of $\unit[2.6]{eV}$ while the direct gap is 
$\unit[2.8]{eV}$, when the experimental lattice constant of
$\unit[3.16]{\text{\AA}}$ is used.
The question of how well the one-shot G$_0$W$_0$@LDA approximation describes 
the true electronic structure of this system remains open. Partially 
self-consistent GW$_0$\cite{Shi_PRB2013} and fully self-consistent 
GW\cite{Chei_PRB2012} calculations have been shown to consistently yield direct 
band gaps of $\unit[2.75]{}-\unit[2.80]{eV}$.

\section{Static screening}\label{sec:epsilon}
In this section we present a detailed investigation of the (static)
dielectric properties of monolayer MoS$_2$. This serves a dual
purpose. First, it illustrates the origin of the slow convergence of
the GW results presented in the previous section (and the BSE results
presented in the next section). Secondly, it shows that the usual definition
of the macroscopic dielectric constant of a periodic solid is not meaningful
when applied to a 2D system represented in a periodic supercell.
We discuss the difference between screening in 2D and 3D which becomes
particularly pronounced in the $q\to 0$ limit with large consequences for
the calculation of optical excitations with static screening of
the electron-hole interaction (see next section).

\subsection{3D macroscopic dielectric constant}\label{sec:eps3D}
The microscopic dielectric function determines the relation between a weak 
external potential and the total potential in the material, 
\begin{equation}
V_{\text{tot}}(\mathbf{r}) = \int \! d\mathbf{r}' \,
\epsilon^{-1}(\mathbf{r}, \mathbf{r}') V_{\text{ext}}(\mathbf{r}').
\end{equation}
For a periodic system the dielectric function can be conveniently expressed in 
plane waves 
\begin{equation}\label{eq:epsreci}
\epsilon^{-1}(\mathbf{r}, \mathbf{r}') =
\sum_{\mathbf G \mathbf G'}\sum_{\mathbf q}
e^{i(\mathbf G + \mathbf q)\mathbf r}
\epsilon^{-1}_{\mathbf G \mathbf G'}(\mathbf q)
e^{-i(\mathbf G' + \mathbf q)\mathbf r'},
\end{equation}
where $\mathbf G$ is a reciprocal lattice vector, $\mathbf q$ belongs to the 1. 
BZ. Within the random phase approximation (RPA) we have  
\begin{equation}\label{eq:eps}
\epsilon_{\mathbf{G}\mathbf{G}'}(\mathbf{q},\omega) = 
\delta_{\mathbf{G}\mathbf{G}'} - v_c(\mathbf{q}+\mathbf{G})
\chi^0_{\mathbf{G}\mathbf{G}'}(\mathbf{q},\omega),
\end{equation}
where $\chi^0$ is the non-interacting density response function. Here, $v_c$
can be the Fourier representation of either the full or the truncated Coulomb
interaction. For the calculations in  this section we have used a
$\unit[50]{eV}$ cut-off for the reciprocal lattice vectors to account for
local field effects. The non-interacting response function, $\chi^0$, was
constructed from local density approximation (LDA) wave functions and energies
including states up to $\unit[50]{eV}$ above the Fermi level. All calculations
were performed with the projector augmented wave method code GPAW. Details on
the implementation of the dielectric function in the GPAW code can be found in
Ref.~\onlinecite{Yan_PRB2011}.

It follows from Eq. (\ref{eq:epsreci}) that the total potential resulting from 
a plane wave external potential $V_0e^{i \mathbf{q}\cdot\mathbf{r}}$ has the 
form
\begin{align}\label{periodic}
 V_{\text{tot}}(\mathbf{r}) =
\widetilde V_{\mathbf{q}}(\mathbf{r}) e^{i \mathbf{q} \cdot \mathbf{r}}
\end{align}
where $\widetilde V_{\mathbf{q}}(\mathbf{r})$ is a lattice periodic function. 
We thus define the macroscopic dielectric constant as
\begin{align}\label{epsmacro}
\frac{1}{\epsilon_M(\mathbf{q})} \equiv
\frac{\langle\widetilde V_{\mathbf{q}} \rangle_\Omega}{V_0} =
\epsilon^{-1}_{\mathbf{0}\mathbf{0}}(\mathbf{q}),
\end{align}
where $\langle\ldots\rangle_\Omega$ denotes a spatial average over a unit cell. 
Note that in general $\epsilon_M(\mathbf{q},\omega) \neq 
\epsilon_{\mathbf{0}\mathbf{0}}(\mathbf{q},\omega)$ because of local field 
effects.\cite{Adler_PhysRev1962, Wiser_PhysRev1963}

\begin{figure}[t]
\begin{centering}
\includegraphics[width=0.85\columnwidth,clip=]{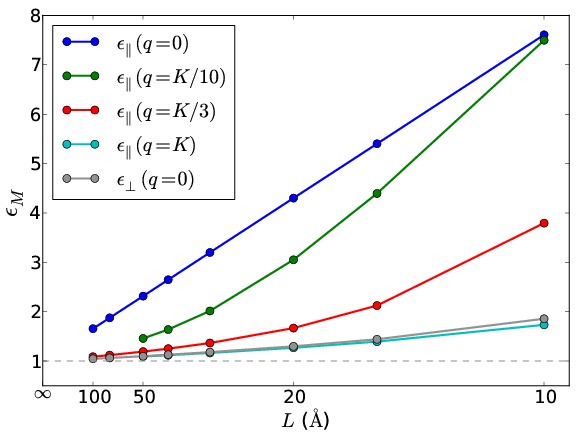}
\par\end{centering}
\caption{(Color online). The 3D static macroscopic dielectric constant 
$1/\epsilon_{\mathbf{0}\mathbf{0}}^{-1}(\mathbf q)$ of monolayer MoS$_2$ as a 
function of the interlayer separation, $L$. $\epsilon_\Vert$ 
is the dielectric constant with polarization parallel to the monolayer and 
$\epsilon_\bot$ is the dielectric constant for polarization orthogonal to 
the layer.}
\label{fig:dielectric}
\end{figure}

To explicitly demonstrate that Eq. (\ref{epsmacro}) does not apply to
low-dimensional materials, we have calculated the macroscopic
dielectric constant as a function of the layer separation, $L$.
The results are shown in Fig. \ref{fig:dielectric} for
different values of the in-plane momentum transfer $q$. We also show the 
dielectric constant corresponding to polarization orthogonal to the layer. 
Clearly the macroscopic dielectric constant approaches unity for all
$q$-vectors in the limit of large interlayer separation.
This occurs because the total field is
averaged over an increasingly larger vacuum region. 

Previously reported values for the macroscopic dielectric constant of
monolayer MoS$_2$ lie in the range $\unit[4]{}-\unit[8]{}$.
\cite{Chei_PRB2012, Rama_PRB2012, Wirtz_PRB2011}
In these calculations the MoS$_2$ layers were separated by
$\unit[10]{}-\unit[20]{\text{\AA}}$ vacuum. As can be seen
from $\epsilon_\Vert (q=0)$ in Fig. \ref{fig:dielectric} this is
consistent with our results. However, it should also be clear that
numbers depend on the distance between layers and in fact are not
meaningful.

\subsection{2D macroscopic dielectric constant}\label{sec:eps2D}
For a 2D material, the average of the total potential in the definition of the 
macroscopic dielectric constant must be confined to the region of the material. 
Since Eq. (\ref{periodic}) still holds for a 2D material when $\mathbf q$ is
confined to the plane of the material, we average
the in-plane coordinates ($\mathbf r_\Vert$) over the unit cell area
$A$ and the out-of-plane coordinate ($z$) from $z_0-d/2$ to $z_0+d/2$
where $z_0$ denotes the center of the material and $d$ its width.  The
2D macroscopic dielectric constant then becomes
\begin{align}\label{eps_2d}
\frac{1}{\epsilon^{2D}_{M}(\mathbf{q_\Vert})}
&\equiv\frac{\langle\widetilde V_{\mathbf{q}}\rangle_{A,d}}{V_0}\notag\\
& = \frac{2}{d}\sum_{G_\bot} e^{iG_\bot z_0} \frac{\sin(G_\bot d/2)}{G_\bot}
\epsilon^{-1}_{\mathbf{G}\mathbf{0}}(\mathbf{q_\Vert}),
\end{align}
where the sum is over all $\mathbf G$ with $\mathbf G_\Vert = \mathbf 0$.
In this work we have taken $d=\unit[6.15]{\text{\AA}}$ corresponding to the
interlayer separation in bulk MoS$_2$. We shall return to the problem of
chosing $d$ below.

\begin{figure}[t]
\begin{centering}
\includegraphics[width=0.85\columnwidth,clip=]{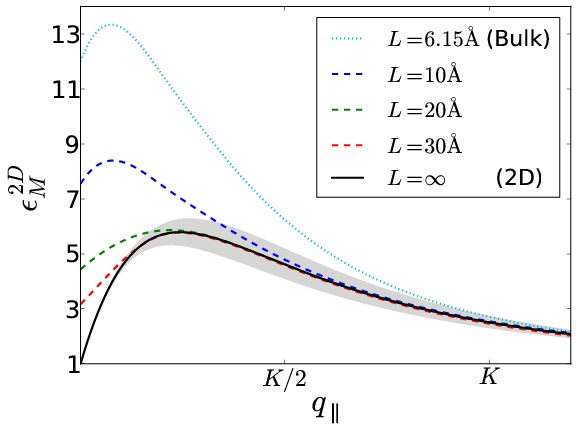}
\par\end{centering}
\caption{(Color online). Static macroscopic dielectric constant for a single 
layer of MoS$_2$ calculated along the $\Gamma$-K line. The calculations are 
performed using Eq. \eqref{eps_2d} with the microscopic dielectric constant, 
$\epsilon^{-1}_{\mathbf{G}\mathbf{G}'}(\mathbf{q})$, evaluated from Eq. 
(\ref{eq:eps}) with either the bare Coulomb interaction (dotted and dashed 
lines) or truncated Coulomb interaction (full black line). The grey area 
represents the result obtained when the averaging region perpendicular to the 
layer, $d$, is varied by $\pm 10\%$. The dotted line corresponds to a layer 
separation of $\unit[6.15]{\text{\AA}}$ and thus coincide with the dielectric
constant of bulk MoS$_2$. The curves have been interpolated from a
$32\times 32$ $q$-point mesh.}
\label{fig:dielectric2D}
\end{figure}

The results for the static dielectric constant evaluated from Eq.
(\ref{eps_2d}) using the bare Coulomb interaction is shown in Fig.
\ref{fig:dielectric2D} for four different layer separations.
The result for $L = d=\unit[6.15]{\text{\AA}}$ coincides with the
3D dielectric constant of  bulk MoS$_2$ given by Eq. (\ref{epsmacro}).
The result obtained with the truncated Coulomb interaction is shown in black; it
represents the case of infinite layer separation. Before discussing the results,
it is instructive to consider the potential arising from a 2D charge density 
fluctuation of the form,
\begin{equation}\label{eq:dipole}
n(\mathbf{r}) =
n_0 e^{i\mathbf q_\Vert \cdot \mathbf r_\Vert} \delta(z),
\end{equation}
The corresponding potential follows from Poisson's equation\footnote{This is 
most easily seen by performing a 3D Fourier transformation of $\delta n$, then 
multiplying by $1/q^2$ and Fourier transforming back to real space.}
\begin{equation}\label{eq:phi_ind}
\phi(\mathbf{r}) =\frac{n_0}{q_\Vert} e^{-i\mathbf q_\Vert \cdot  \mathbf 
r_\Vert} e^{-q_\Vert |z|}.
\end{equation}
It follows that the potential perpendicular to the layer falls off 
exponentially over a characteristic distance of $1/q_\Vert$. This explains why 
in general $\epsilon_M^{2D}(\mathbf q_\Vert)$ coincides with the isolated layer 
result for $q_\Vert \gtrsim 1/L$. 

The variation of $\epsilon_M^{2D}$ when the parameter $d$ is changed
by $\pm 10\%$ is indicated by the shaded region in Fig.
\ref{fig:dielectric2D}. To the left of the maximum,
$\epsilon_M^{2D}(\mathbf q_\Vert)$ is insensitive to $d$ since the
induced potential is more or less constant over the averaging region.
To the right of the maximum, the variation in $\epsilon_M^{2D}(\mathbf
q_\Vert)$ follows the $\pm 10 \%$ variation in $d$. This is because
for these wave vectors the induced potential has essentially vanished
at the borders of the averaging region. In general, increasing
(decreasing) $d$ will decrease (increase) $\epsilon_M^{2D}(\mathbf
q_\Vert)$ in the large wave vector region. 

Another characteristic feature of the potential in Eq. (\ref{eq:phi_ind})
is the $1/q_\Vert$ scaling which should be compared with the $1/q^2$
form of the Coulomb potential from a 3D charge oscillation. Since the
non-interacting response function, $\chi_{\mathbf{0}\mathbf{0}}^0(\mathbf{q})$,
scales as $\sim q^2$ for $q \rightarrow 0$ for both 2D and 3D systems, it 
follows from Eq. (\ref{eq:eps}) that $\epsilon_M^{2D}(0)=1$,
while this is in general not the case in 3D.
In our calculations, the effect of interlayer interactions is eliminated
by using a truncated Coulomb interaction of the form
$v_c(\mathbf r)=(1/r)\theta(R_c-z)$. 
For $q_z=0$ and in the limit of small $q_\Vert$, the Fourier representation of 
the truncated Coulomb interaction becomes
$v^{2D}(\mathbf q) = \frac{4\pi R_c}{|\mathbf q|}$,
i.e. it scales as $1/q$ as the potential from the 2D charge density 
wave ensuring the correct limit $\epsilon_M^{2D}(0)=1$.

Finally, we note that previous studies\cite{Cudazzo_PRB2011,
Berkelbach_PRB2013} have employed a strict 2D model for the dielectric function
in the small $q$ limit of the form $\epsilon(q_\Vert)=1+\alpha q_\Vert$. This
form is convenient as it leads to an analytical expression for the screened
interaction in 2D.\cite{Keldysh_JETP1979} Our definition differs by being a 3D
(or quasi 2D) quantity valid for general $q_\Vert$.

\subsection{Screened interaction}\label{subsec:screenedinteraction}
\begin{figure}[t]
\begin{centering}
\includegraphics[width=0.85\columnwidth,clip=]{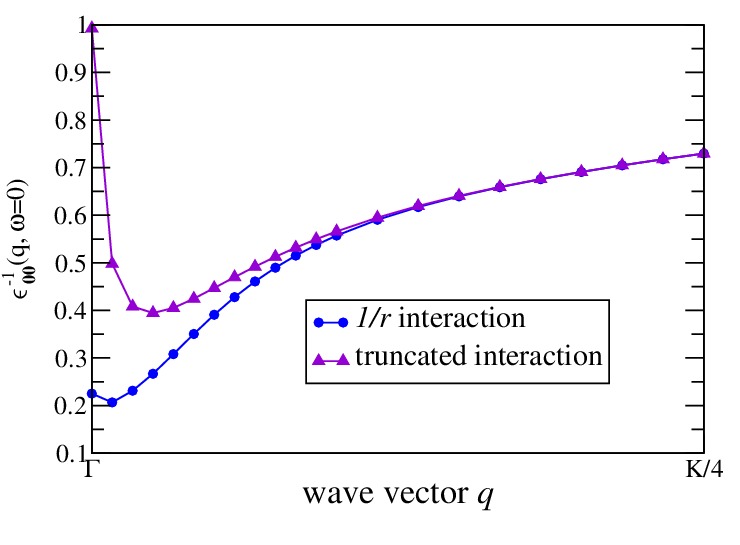}
\par\end{centering}
\caption{(Color online). The 3D static inverse dielectric constant 
$\epsilon_{\mathbf{0}\mathbf{0}}^{-1}(\mathbf{q})$ of monolayer MoS$_2$
calculated in the RPA for different values of in-plane momentum transfer
$q$ along the $\Gamma$-K direction.
The separation between layers is $L=\unit[20]{\text{\AA}}$.
Note that neither of the quantities can be interpreted as the macroscopic
dielectric constant of the monolayer (this quantity is the black curve in Fig.
\ref{fig:dielectric2D}).}
\label{fig:dielectric_q}
\end{figure}

In Fig. \ref{fig:dielectric_q} we show
$\epsilon_{\mathbf{0}\mathbf{0}}^{-1}$ as a function of $q_{\Vert}$ 
evaluated with and without the truncated Coulomb interaction. For small $q$,
the two curves differ significantly due to the long range nature of the
induced potential (\ref{eq:phi_ind}).
At large $q$ ($\sim K/2$), the induced potential decays 
within the cutoff range for the truncated Coulomb interaction and therefore no 
difference can be seen between the two methods.
We emphasize that neither of the dielectric constants shown in the figure can 
be interpreted as a dielectric constant of monolayer MoS$_2$, since they give 
the average potential over the supercell and not over the MoS$_2$ layer. In 
particular their value will be highly dependent on the size of the unit cell 
(in the limit of infinite layer separation both will equal 1 for all $q$). 
Nevertheless, this quantity is a crucial ingredient of both the GW self-energy 
and the BSE kernel as it provides the screening of the divergent term of the 
Coulomb interaction.

For $q=0$ the Coulomb kernel diverges and we approximate $W(q=0)$ by the 
integral
\begin{align}
W_{\mathbf{0}\mathbf{0}}(\mathbf{q}=0) &=
\frac{1}{\Omega_\Gamma} \int_{\Omega_\Gamma} \! d\mathbf{q} \,
v_c(\mathbf{q})\epsilon^{-1}_{\mathbf{0}\mathbf{0}}(\mathbf{q}) \notag\\
&\approx\frac{1}{\Omega_\Gamma}\epsilon^{-1}_{\mathbf{0}\mathbf{0}}(\mathbf{q}=0)
\int_{\Omega_\Gamma} \! d\mathbf{q} \, v_c(\mathbf{q}),
\label{eq:W_int}
\end{align}
where $\Omega_{BZ}$ is the Brillouin zone volume and $\Omega_\Gamma$ is a small 
volume containing $\mathbf{q}=0$. In isotropic systems 
$\epsilon^{-1}_{\mathbf{0}\mathbf{0}}(\mathbf{q})$ is usually constant
in the vicinity of $\mathbf{q}=0$ and the approximation works well.
However, when $\epsilon^{-1}$ is evaluated with the truncated
Coulomb interaction, $\epsilon^{-1}_{\mathbf{0}\mathbf{0}}$ acquires
much more structure for small $q$ as can be seen from
Fig. \ref{fig:dielectric_q}. Thus, for coarse $k$-point samplings we 
will underestimate the $\Gamma$-point screening since we simply use 
$\epsilon^{-1}_{\mathbf{0}\mathbf{0}}(\mathbf{q}=0) = 1$.

The linear behavior of the screened interaction for small $q$ suggests
that a better approximation for $W_{\mathbf{0}\mathbf{0}}(\mathbf{q}=0)$ would
be
\begin{equation}
 W_{\mathbf{0}\mathbf{0}}(\mathbf{q}=0) = \frac{1}{\Omega_\Gamma}
\int_{\Omega_\Gamma} \! 
d\mathbf{q} \,
v_c(\mathbf{q}) \Big[1 + \mathbf{q} \cdot \nabla_\mathbf{q}
\epsilon_{\mathbf{0}\mathbf{0}}^{-1}(\mathbf{q}) \Big|_{\mathbf{q}=0}\Big].
\end{equation}
Since the dielectric matrix in RPA is 
$\epsilon_{\mathbf{G}\mathbf{G}'}(\mathbf{q})=1-v^c(\mathbf{q})\chi^0_{
\mathbf{G}\mathbf{G}'}(\mathbf{q})$, we can derive an analytic expression for 
the first order Taylor expansion in $q$, and its inverse. These quantities can 
be evaluated with vanishing additional cost, but we will leave the assessment 
of this approximation to future work.

\section{Optical absorption spectrum}\label{sec:BSE}
In this section we present many-body calculations of the optical
absorption spectrum of monolayer MoS$_2$ by solving the Bethe-Salpeter
Equation (BSE) under the standard assumption of static screening of
the electron-hole interaction. As for the GW band gap, we find that
the use of a truncated Coulomb interaction is essential to avoid
interlayer screening and obtain well converged exciton binding
energies. Furthermore, the very strong q-dependence of the 2D static
dielectric function around $q=0$, leads to very slow $k$-point
convergence for the exciton binding energy.

In order to obtain an accurate absorption spectrum including excitonic effects 
we calculate the response function from the Bethe-Salpeter Equation (BSE).
Using 
the standard assumption of a static dielectric screening of the electron-hole 
(e-h) interaction, the BSE\cite{BSE} can be recast as an effective two-particle
Hamiltonian,\cite{Strinati_PRB1984} which is diagonalized on a
basis of electron-hole pairs. In this way the excitonic eigenstates can be 
expressed as a linear combination of single-particle transitions
\begin{align}\label{state}
|\lambda\rangle=\sum_{vck}A^\lambda_{vck}|vck\rangle,
\end{align}
where $v$, $c$, and $k$ denote valence band, conduction band and Brillouin zone 
wave vector, respectively. The absorption spectrum is proportional to the 
imaginary part of the macroscopic dielectric function, which in the 
Tamm-Dancoff approximation can be written
\begin{align}\label{eps_2}
\epsilon_2(\omega)=&2\pi\lim_{\mathbf{q}\rightarrow 
0}v_c(\mathbf{q})\sum_\lambda\delta(\omega-E_\lambda)\notag\\
&\times\bigg|\sum_{vc\mathbf{k}}A^\lambda_{vck}\langle 
v\mathbf{k}-\mathbf{q}|e^{-i\mathbf{q}\cdot\mathbf{r}}|c\mathbf{k}
\rangle\bigg|^2,
\end{align}
where $E_\lambda$ are the eigenvalues associated with $|\lambda\rangle$.

In all calculations we have included a single valence band and a single 
conduction band in the BSE Hamiltonian. We have tested that the first excitonic 
peak is completely unaffected if we instead include 6 valence bands and 4 
conduction bands. This is also expected since the highest (lowest) 
valence (conduction) band is well isolated from the remaining bands at K 
where the exciton is centered, see Fig. \ref{fig:bandstructure}.
For the screening we have included 65 bands in the evaluation of the
response function, which is sufficient for converged results.
Increasing the number of bands to 300 affects the position of the first
exciton by less than $\unit[10]{meV}$. The plane wave cutoff for the
response function (local field effects) was set to $\unit[50]{eV}$
and we checked that the excitonic binding energy changed by less than
$\unit[10]{meV}$ when increasing the cutoff to $\unit[200]{eV}$. The dependence
on $k$-point sampling and interlayer separation will be examined below. Details
on the implementation of the BSE method in the GPAW code can be found in Ref.
\onlinecite{Yan_PRB2012}.

\subsection{Convergence tests}\label{sec:BSEconvergence}
\begin{figure}[t]
\begin{centering}
\includegraphics[width=0.85\columnwidth,clip=]{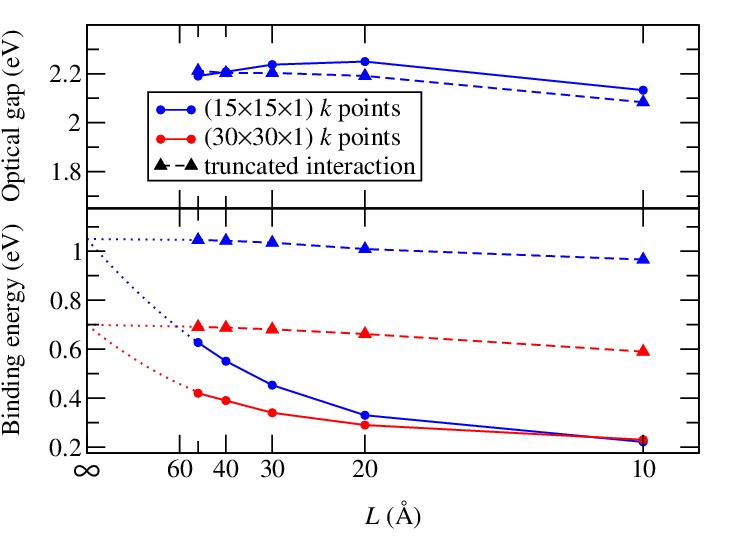}
\par\end{centering}
\caption{(Color online). Optical gap and binding energy of the lowest exciton
in monolayer MoS$_2$ as a function of interlayer separation calculated
from the BSE and the G$_0$W$_0$ quasiparticle gap. Results with the full
$1/r$ Coulomb interaction (full lines) and the truncated interaction
(dashed lines) are shown. Dotted lines give an estimation for extrapolation
to infinite $L$.}
\label{fig:bse_vacuum}
\end{figure}

In the lower panel of Fig. \ref{fig:bse_vacuum}, we show the exciton binding 
energy as a function of interlayer separation calculated for different 
$k$-point samplings using the bare Coulomb interaction and the truncated
Coulomb interaction. With the bare Coulomb interaction, the obtained
results are far from convergence, even for $L = \unit[50]{\text{\AA}}$.
The dependencies on the layer separation and number of $k$ points
is very similar as for the quasiparticle gap discussed in Sec.
\ref{subsec:GWresults}, even on a quantitative level. Therefore, the optical
gap, which is given by the difference of the QP gap and the exciton binding
energy, is almost indenpedent of $L$ and whether or not the truncation method
is used, as shown in the upper panel. This is consistent with the
observations in Ref. \onlinecite{Komsa_PRB2012}.

\begin{figure}[t]
\begin{centering}
\includegraphics[width=0.85\columnwidth,clip=]{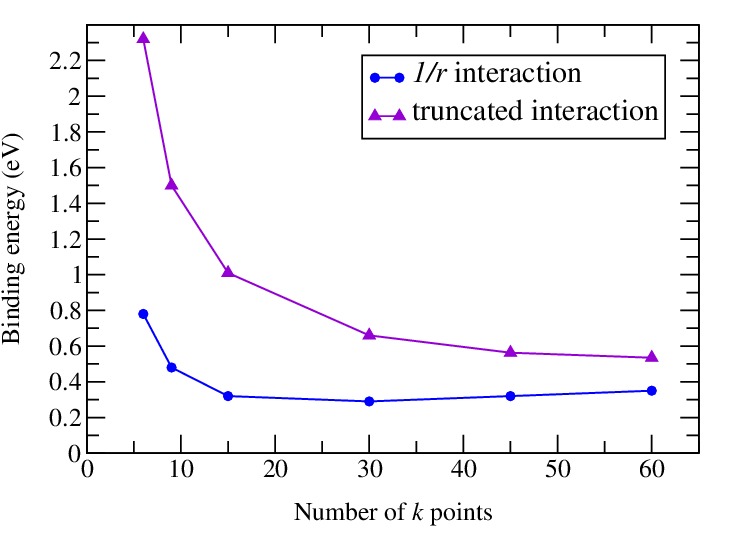}
\par\end{centering}
\caption{(Color online). Binding energy of the lowest exciton in monolayer 
MoS$_2$ as a function of $k$-point sampling for a supercell with a
layer separation of $L=\unit[20]{\text{\AA}}$.}
\label{fig:bse_kpoints}
\end{figure}

The convergence of the binding energy with respect to the $k$-point sampling
is plotted in Fig. \ref{fig:bse_kpoints} for an interlayer separation of 
$\unit[20]{\text{\AA}}$. The truncated Coulomb kernel gives a much slower
convergence with respect to the number of $k$-points 
than the bare Coulomb interaction. However, it should be clear from Fig. 
\ref{fig:bse_vacuum} that the binding energy obtained with the bare Coulomb 
interaction converges to a value which is highly dependent on the interlayer 
separation. In Ref. \onlinecite{Molina_PRB2013}, convergence was found
with $18\times18$ $k$ points, but for a layer separation of only
$\unit[24]{\text{\AA}}$. The obtained exciton binding energy was around
$\unit[0.2]{eV}$. According to our results, this is much too weak due to
interlayer screening.

The slow $k$-point convergence observed when using the truncated Coulomb 
interaction is related to the $q$-dependence of the screening 
in two-dimensional systems. As demonstrated by Eq. (\ref{eq:W_int}) and Fig. 
\ref{fig:dielectric_q} (blue curve), a too low $k$-point sampling leads to an 
underestimation of the screening in the vicinity of $q=0$ and thus an 
overestimation of the exciton binding energy.

\subsection{Results}\label{sec:BSEresults}
\begin{figure}[t]
\begin{centering}
\includegraphics[width=0.85\columnwidth,clip=]{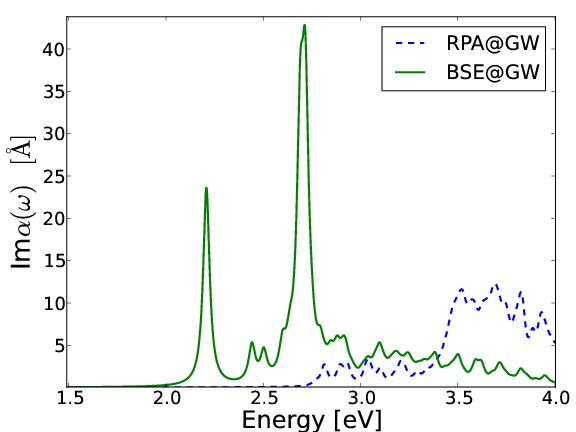}
\par\end{centering}
\caption{(Color online). Absorption spectrum of single layer MoS$_2$ calculated 
with the RPA and BSE using the G$_0$W$_0$ quasiparticle band structure. The 
calculation has been performed with a truncated Coulomb interaction to avoid 
interactions between repeated layers and with a $45\times 45$ $k$-point grid.}
\label{fig:abs}
\end{figure}

From the convergence tests described above we conclude that the BSE calculations
are (nearly) converged if we use a truncated Coulomb interaction and a 
$45\times45$ $k$-point sampling. With these setting we have calculated the BSE
spectrum on top of a G$_0$W$_0$ quasiparticle band structure obtained with the
same parameters. The resulting absorption spectrum is shown in 
Fig. \ref{fig:abs}. We also show an RPA calculation, i.e. neglecting 
electron-hole interactions in the BSE, performed on top of the same G$_0$W$_0$ 
band structure for comparison. With electron-hole interaction included, we 
obtain an exciton binding energy of $\sim \unit[0.6]{eV}$, whereas RPA does
not show an exciton peak and simply gives an absorption edge at the band gap.

Experimentally, the absorption spectrum of single layer MoS$_2$ exhibits a 
spin-orbit split peak around $\unit[1.9]{eV}$.\cite{Mak_PRL2010} Since we have 
not included  spin-orbit coupling in our calculations, the spectrum
Fig. \ref{fig:abs} only shows a single peak at low energies.
However, it has previously been shown\cite{Rama_PRB2012, Shi_PRB2013}
that the spin-orbit coupling does not have a large effect on the exciton
binding energy and only results in a split excitonic peak.
The main peak in the BSE@G$_0$W$_0$ spectrum is
situated at $\unit[2.2]{eV}$ which is $\unit[0.3]{eV}$ 
higher than the experimental value. At present we cannot say if this is due to 
an insufficient description of the quasiparticle gap within G$_0$W$_0$ or 
underestimation of the exciton binding energy in BSE with a static 
electron-hole interaction.

From the above discussion it should be clear that it is extremely 
challenging to converge the exciton binding energy with respect to interlayer 
separation and $k$ points. In general, the optical gap is much easier to 
converge with respect to interlayer separation than either the quasiparticle 
gap or the exciton binding energy.\cite{Wirtz_PRL2006, Komsa_PRB2012}
Nevertheless, for many physical applications it is of importance to
obtain accurate values for both the quasiparticle gap and the exciton
binding energy separately. In Ref. \onlinecite{Komsa_PRB2012}
the exciton binding energy was obtained by $1/L$ extrapolation of
the quasiparticle gap calculated in a range of interlayer separations
between $\unit[10]{}$ and $\unit[20]{\text{\AA}}$
and assuming the same dependence for the exciton binding energy. Our results 
indicate that one should be cautious with such extrapolations. This is because 
the screening at different $q$-points has a very different dependence 
on interlayer separation, which results in different 
convergence behavior at different $k$-point samplings (see Fig. 
\ref{fig:bse_vacuum} full lines). The extrapolation 
procedure may therefore not give reliable results, since higher $k$-point 
samplings are required at larger interlayer separation. We are aware that the 
convergence issues may depend a lot on the implementation of the BSE method. 
However, we have previously performed the same calculations with 
YAMBO\cite{YAMBO} code, which produced very similar convergence behavior for 
quasiparticle gap and exciton binding energy (also using truncated
Coulomb cutoff and $45\times45$ $k$-point sampling).

\section{Conclusions}\label{sec:conclusion}
We have presented a careful investigation of the dielectric
properties, band gap and excitonic states in a two-dimensional
semiconductor exemplified by monolayer MoS$_2$.  We have demonstrated
that the "standard" macroscopic dielectric constant used for solids is
not applicable (meaningless) to supercells describing the 2D material
as an infinite array of parallel sheets, and therefore replaced it
by a 2D version in which the induced field is averaged over
the extent of the material rather than over the entire supercell.
We showed that the effect of interlayer screening leads to
underestimation of the band gap and exciton binding energy by up to more than
$\unit[0.5]{eV}$ for layer separations $< \unit[30]{\text{\AA}}$.
The reason for this is that interlayer screening affects
$\epsilon(q)$ for $q<1/L$ where $L$ is the distance
between layers in the supercell. Since it is the small $q$ limit
of $\epsilon(q)$ that is most important for the screened
interaction $W(q)=\epsilon^{-1}(q)/q^2$, the effect cannot be
neglected. Here we have circumvented the problem by using a
truncated Coulomb interaction that explicitly cuts off the interaction
between neighboring layers.

The properly defined 2D dielectric function $\epsilon_M^{2D}(q)$ has a
very sharp wave vector dependence for small $q$ and satisfies
$\epsilon_M^{2D}(0)=1$ in general. This has the consequence that
quasiparticle- and optical excitations obtained from the GW and
Bethe-Salpeter Equation, respectively, require much denser $k$-point
grids than experience from 3D systems would suggest. For MoS$_2$ we
find that a precision of $\unit[0.2]{eV}$ requires $k$-point grids of at least
$30\times 30$. Interestingly, the effect of interlayer screening and too
small $k$-point grids have opposite effects on the band gap and exciton
energies leading to fortuitous error cancellation. Our calculations applying
the truncated Coulomb interaction and $45\times 45$ $k$ points give
G$_0$W$_0$@LDA gaps of $\unit[2.77]{eV}$ (direct) and $\unit[2.58]{eV}$
(indirect) and binding energy of the lowest exciton of $\unit[0.55]{eV}$.
This places the lowest exciton at $\sim \unit[2.2]{eV}$ which is
$\unit[0.3]{eV}$ higher than the experimental result. This difference
may be due to the G$_0$W$_0$@LDA approximation or the use of static
screening in the BSE.

\section{Acknowledgement}
KST acknowledges support from the Danish Council for Independent Research's 
Sapere Aude Program through grant no. 11-1051390. The Center for Nanostructured
Graphene (CNG) is sponsored by the Danish National Research Foundation, Project 
DNRF58.

\bibliography{bibfile}{}

\end{document}